%% file: 3C286-pol-calib-v7.0.tex
\documentclass[structabstract]{aa}  

\usepackage{natbib}
\usepackage{txfonts}
\usepackage{graphicx}

\begin{document}

\title{3C~286: a bright, compact, stable, and highly polarized calibrator for millimeter-wavelength observations}

\titlerunning{3C~286: a polarization calibrator for millimeter observations}

\author{Iv\'{a}n Agudo\inst{1,2}
            \and
            Clemens Thum\inst{3} 
            \and  
            Helmut Wiesemeyer\inst{4,5}
            \and
            Sol N. Molina\inst{1}
            \and 
            Carolina Casadio\inst{1} 
            \and  
            Jos\'{e} L. G\'{o}mez\inst{1}
            \and
            Dimitrios Emmanoulopoulos\inst{6}
            }

\authorrunning{Agudo et~al.}
\institute{Instituto de Astrof\'{i}sica de Andaluc\'{i}a, CSIC, Apartado 3004, 18080, Granada, Spain\\
               \email{iagudo@iaa.es}
               \and
               Institute for Astrophysical Research, Boston University, 725 Commonwealth Avenue, Boston, MA 02215, USA
               \and
               Institut de Radio Astronomie Millim\'{e}trique, 300 Rue de la Piscine, 38406 St. Martin d'H\`{e}res, France
               \and
               Max-Planck-Institut f\"ur Radioastronomie, Auf dem H\"ugel, 69, D-53121, Bonn, Germany
               \and
               Instituto de Radio Astronom\'{i}a Milim\'{e}trica, Avenida Divina Pastora, 7, Local 20, E--18012 Granada, Spain
               \and
               School of Physics and Astronomy, University of Southampton, Southampton SO17 1BJ, UK
               }


\abstract
{A number of millimeter and submillimeter facilities with linear polarization observing capabilities have started operating during last years. These facilities, as well as other previous millimeter telescopes and interferometers, require bright and stable linear polarization calibrators to calibrate new instruments and to monitor their instrumental polarization. The current limited number of adequate calibrators implies difficulties in the acquisition of these calibration observations.}
{Looking for additional linear polarization calibrators in the millimeter spectral range, in mid-2006 we started monitoring \object{3C~286}, a standard and highly stable polarization calibrator for radio observations.}
{Here we present the 3 and 1\,mm monitoring observations obtained between September 2006 and {January 2012} with the XPOL polarimeter on the IRAM 30\,m Millimeter Telescope.}
{Our observations show that \object{3C~286} is a bright source of constant total flux with 3\,mm flux density $S_{\rm{3mm}}=(0.91\pm0.02)$\,Jy.
The  3\,mm linear polarization degree ($p_{\rm{3mm}}=[13.5\pm0.3]$\,\%) and polarization angle ($\chi_{\rm{3mm}}=[37.3\pm0.8]^{\circ}$, expressed in the equatorial coordinate system) are also constant during the time span of our observations.
Although with poorer time sampling and signal-to-noise ratio, our 1\,mm observations of \object{3C~286} are also reproduced by a constant source of 1\,mm flux density ($S_{\rm{1mm}}=[0.30\pm0.03]$\,Jy), polarization fraction ($p_{\rm{1mm}}=[14.4\pm1.8]$\,\%), and polarization angle ($\chi_{\rm{1mm}}=[33.1\pm5.7]^{\circ}$).}
{This, together with the previously known compact structure of \object{3C~286} --extended by $\sim3.5^{\prime\prime}$ in the sky-- allow us to propose \object{3C~286}  as a new calibrator for both single dish and interferometric polarization observations at 3\,mm, and possibly at shorter wavelengths.}


\maketitle

\section{Introduction}
\label{intr}
Despite the recent start of operations of a considerable number of millimeter observing facilities with ability for high precision polarization measurements (e.g.,  ALMA, PLANK, IRAM 30\,m Telescope and PdB Interferometer, SMA, KVN, BOOMERANG, CBI, MAXIPOL, QUaD, WMAP, and BICEP), the number of suitable polarization calibrators is still limited.
The most commonly used sources for calibration of millimeter polarization observations are the \object{Crab} nebula \citep[\object{Taurus A},][]{Aumont:2010p12769}, \object{Centaurus A}  \citep{Zemcov:2010p16834}, the limb of the \object{Moon} \citep{Barvainis:1988p16835,Thum:2003p17117}, and the diffuse Galactic emission \citep{Matsumura:2010p16833}.
An ideal linearly polarized calibrator is a point-like, bright and constant source of high linear-polarization fraction with well defined and constant polarization angle. 
For most facilities designed to study the polarization properties of the cosmic microwave background, which still do not aim at observations with angular resolutions much better than $\sim5^{\prime}$, one or more among the four calibrators mentioned above fulfill most of these conditions.
However, the relatively large angular extension of all these four polarization calibrators \citep[$\gtrsim5^{\prime}$, where $5^{\prime}$ is the approximate angular extension of the Crab nebula at millimeter wavelengths;][]{Aumont:2010p12769} may make difficult polarization calibration observations at angular resolutions better than that.
Even when this is not the case, the poor coverage of adequate calibrators in the sky often reveals the need for new polarized sources suitable for calibration.

In this paper we present the results from series of millimeter measurements with the IRAM 30\,m Telescope to characterize \object{3C~286} (with J2000 equatorial coordinates: $\alpha=13^{\rm{h}}$ $31^{\rm{m}}$ $8.3^{\rm{s}}$, $\delta=+30^{\circ}$ 30$^{\prime}$ 33$^{\prime\prime}$) as a new bright, compact, stable, and highly polarized calibrator for short millimeter observations.
This source has been widely used for calibration at centimeter radio wavelengths both for total flux \citep{Baars:1977p16842,Ott:1994p16851,Kraus:1999p13882} and linear polarization observations \citep{Perley:1982p6054,McKinnon:1992p16861,Taylor:2001p16867,2002.VLBA.SM.30}.

\object{3C~286} is a bright compact steep spectrum radio quasar \citep{Peacock:1982p16871,Fanti:1985p16888} at a redshift $z=0.849$ \citep{Burbidge:1969p16881}.
Radio imaging with the Very Large Array at sub-arcsecond resolution has revealed an extended structure composed of three misaligned bright features \citep[e.g.,][]{Akujor:1995p16994}. 
Among these three structures, the second one in brightness is located $2.6^{\prime\prime}$ West--South--West from the brightest one.
Both of them are linked by a nearly straight jet-like bridge of radio emission seen in high dynamic range images. 
The third emission region is located at $0.8^{\prime\prime}$ East from the brightest one. 
Among these three emission features, the two brightest ones dominate the linear polarization emission, and show their electric vector oriented nearly parallel to the axis of the jet like structure between them.
The integrated electric vector polarization angle of the source\footnote{To express polarization angles, we follow throughout this paper the IAU convention, which counts East from North in the equatorial coordinate system.} $\chi\approx33^{\circ}$ at all observing wavelengths from 20\,cm to 7\,mm \citep{Perley:1982p6054,Taylor:2001p16867}, see also {\tt http://www.vla.nrao.edu/astro/calib/polar/}.
The stability of $\chi$ along radio frequencies has also been demonstrated through polarization angle rotation measure studies \citep{Rudnick:1983p17021}.
Very long baseline interferometric radio observations at milliarcsecond resolution of the brighter, and more compact, feature have revealed a well defined jet structure initially oriented in the South--West direction up to scales $>60$\,milliarcseconds, with the linear polarization fraction ranging from $p\approx1$\,\% and $p\approx20$\,\% and $\chi$ oriented nearly (but not exactly) parallel to the local jet axis \citep{Jiang:1996p17007}.
These properties are not uncommon in radio loud quasar jets.
However, \object{3C~286} lacks a highly variable, low polarization, and high brightness temperature ($T_{\rm{b}}\gtrsim10^{10}$--$10^{11}$\,K) radio core \citep{Cotton:1997p17000}, as typical for radio quasars.
This makes \object{3C~286} an excellent total flux and polarization calibrator at radio wavelengths, and provides it with its characteristic non variable, and large integrated linear polarization degree ($p\approx11$\,\%) at radio frequencies ({\tt http://www.vla.nrao.edu/astro/calib/polar/}).
The lack of a typical radio core in \object{3C~286} was interpreted by \citet{Cotton:1997p17000} as produced by strong relativistic de-boosting of the emission from the core when beamed in a direction significantly different from that of the observer's line of sight.

\section{Observations and data reduction}
\label{obs}
\begin{figure}
   \centering
   \includegraphics[width=\columnwidth]{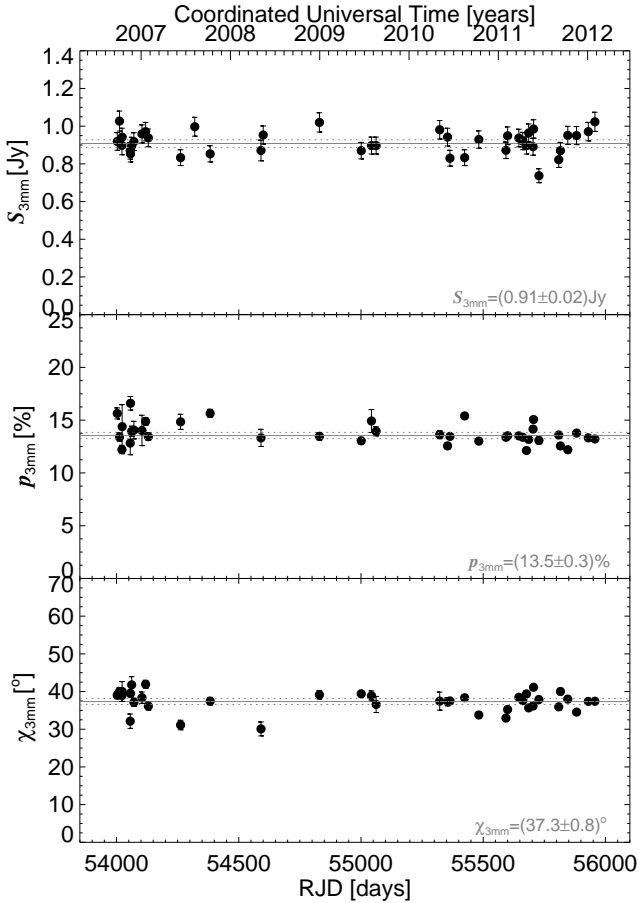}
   \caption{3\,mm total flux and polarization evolution of \object{3C~286}. The fitted constant values of $S_{\rm{3mm}}$, $p_{\rm{3mm}}$, and $\chi_{\rm{3mm}}$ to be used for calibration through observations of \object{3C~286}, and their 95\,\% confidence interval (obtained in Section~\ref{res3mm}), are given on every panel. The horizontal continuous and dotted lines symbolize such constant and 95\,\% interval, respectively.}
   \label{f3mm}
\end{figure}

\begin{figure}
   \centering
   \includegraphics[width=\columnwidth]{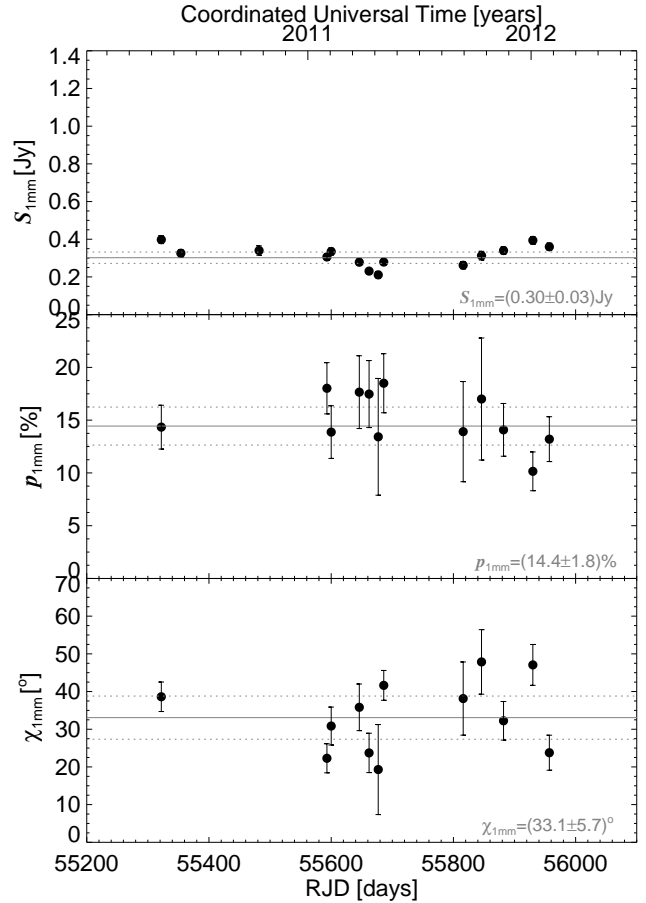}
   \caption{Same as Fig.~\ref{f3mm} but for the 1\,mm data.}
   \label{f1mm}
\end{figure}

The observations presented here were performed with the XPOL polarimeter \citep{2008PASP..120..777T} on the IRAM 30\,m Telescope.
The \object{3C~286} measurements were taken for calibration purposes under the POLAMI (Polarimetric AGN Monitoring at the IRAM-30\,m-Telescope) and the MAPI (Monitoring of AGN with Polarimetry at the IRAM-30\,m-Telescope) programs \citep[e.g.,][]{Agudo:2011p15946}, and for a densely time sampled 3\,mm polarimetric monitoring program of blazar \object{OJ287} \citep{Agudo:2011p14707}.
The time range of \object{3C~286} observations covers from 24 of September, 2006 ($\rm{RJD}=54003$)\footnote{$\rm{RJD,\,i.e.\,reduced\,Julian\,date}=\rm{Julian\,Date}-240000$\,days} to 30 of January, 2012 ($\rm{RJD}=55957$); see Figs.~\ref{f3mm} and \ref{f1mm}, and Tables~\ref{t3mm} and \ref{t1mm}.

\input{./3C286-table-3mm}

\input{./3C286-table-1mm} 

The standard XPOL set-up and calibration scheme introduced in \citet{2008PASP..120..777T} and \citet{Agudo:2010p12104} were used.
Those measurements performed before Spring 2009 made use of the orthogonal linearly polarized A100 and B100 heterodyne receivers tuned at 3\,mm (86\,GHz).
After Spring 2009, we employed the new E090 and E230 pairs of orthogonal linearly polarized receivers at the IRAM 30\,m Telescope to observe simultaneously at 3\,mm and 1\,mm (229\,GHz), where the angular resolution of the telescope (full with at half power) is $28^{\prime\prime}$ and $11^ {\prime\prime}$, respectively.

After a cross-scan pointing of the telescope and an amplitude, and phase calibration measurement \citep[see][]{2008PASP..120..777T}, every XPOL measurement consisting of wobbler switching on-offs was performed for a total integration time $\sim8$\,min.
The signal from every one of the two orthogonal 3\,mm receivers was connected to the XPOL polarimeter, whose output consisted of a 640\,MHz spectrum from each one of such receivers, plus the real and imaginary part of their cross-correlation.
These four XPOL observables are used to compute the 4 Stokes parameters for each measurement \citep{2008PASP..120..777T}.
XPOL hardware limitations only allow to connect 320\,MHz to the remaining bandwidth of the polarimeter. 
This bandwidth was used for the simultaneous 1\,mm observations taken after the end of 2009.

After calibration of the amplitudes and phases from every pair of linearly polarized orthogonal receiver, the instrumental polarization for every observing epoch (estimated through measurements of unpolarized calibrators, i.e. Mars and/or Uranus, when available) was removed from the data.
The instrumental polarization parameters used for 3\,mm observations before Spring 2009 were $Q_{\rm{i}}\approx0.005I$, $U_{\rm{i}}\approx0.003I$, for the contribution of the instrumental polarization to the $Q$ and $U$ Stokes parameters, respectively.
After mid 2009, new instrumental polarization parameters characterized the performance of the new E090 and E230 receiver pairs on the 30\,m Telescope.
These are $Q_{\rm{i}}\approx-0.007I$ and $U_{\rm{i}}\approx-0.003I$ for 3\,mm observations and $Q_{\rm{i}}\approx-0.015I$ and $U_{\rm{i}}\approx-0.015I$ for 1\,mm.
The Jy/K calibration factor ($C_{\rm{Jy/K}}$) for the total flux density scales were computed for every observing epoch before Spring 2009 as in \citet{Agudo:2006p203}.
The resulting $C_{\rm{Jy/K}}$ values were fully consistent with the standard $C_{\rm{Jy/K}}$ calibration factors at 3 and 1\,mm for the IRAM 30\,m Telescope ($6.4$\,Jy/K and $9.3$\,Jy/K, respectively; H. Ungerechts, private communication).
For epochs after mid 2009 we used such standard $C_{\rm{Jy/K}}$ calibration factors.

The total flux density uncertainties given in Tables~\ref{t3mm} and \ref{t1mm} correspond to the statistical uncertainties for every specific measurement plus a 5\,\% systematic factor from the uncertainties in $C_{\rm{Jy/K}}$ \citep{Agudo:2006p203} that was added quadratically.
The linear polarization degree ($p$) and linear polarization angle ($\chi$) uncertainties in Tables~\ref{t3mm} and \ref{t1mm} were estimated from the statistical uncertainties from every measurement, plus a non-systematic contribution computed from the dispersion of $Q$, and $U$ Stokes parameters from measurements of Mars and Uranus.
These dispersion estimates are $\Delta Q_{\rm{i}}=0.003I$, and  $\Delta U_{\rm{i}}=0.002I$, for both 3 and 1\,mm observations. 
These give polarization-uncertainty medians of $\Delta{p}\approx0.5$\,\%,  $\Delta{\chi}\approx1$\,$^{\circ}$ at 3\,mm, and  $\Delta{p}\approx4$\,\%,  $\Delta{\chi}\approx5$\,$^{\circ}$ at 1\,mm for our measurements of \object{3C~286}.
The absolute calibration of the polarization angle of the XPOL polarimeter has a precision of $0.5^{\circ}$ or better \citep{2008PASP..120..777T,Aumont:2010p12769}. 
Hence, we do not consider a significant source of error from such calibration on the uncertainties of our final measurements.

\section{Results and Discussion}
\subsection{3\,mm}
\label{res3mm}
Figure \ref{f3mm} shows the 3\,mm total flux density ($S_{\rm{3mm}}$), linear polarization degree ($p_{\rm{3mm}}$), and polarization angle ($\chi_{\rm{3mm}}$) evolution of \object{3C~286} as measured from mid 2006 to the beginning of 2012 (see also Table~\ref{t3mm}). 
Neither $S_{\rm{3mm}}$, nor $p_{\rm{3mm}}$ and $\chi_{\rm{3mm}}$ show appreciable variability during the time range of our observations.
Indeed, Fig.~\ref{f3mm} resembles the behavior of a non varying total flux and polarization source, as observed for \object{3C~286} at longer centimeter wavelengths.

To verify the stability of $S_{\rm{3mm}}$, $p_{\rm{3mm}}$, and $\chi_{\rm{3mm}}$ with time, we test if these variables can be fitted by a constant. 
Initially we fit a linear model to each one of these data sets.
This model is characterized by a slope ($b$) and an intercept ($a$), i.e. $y=a+bt$, were $t$ is the time and $y$ is either $S_{\rm{3mm}}$, $p_{\rm{3mm}}$, or $\chi_{\rm{3mm}}$. 
The resulting linear-model best-fit parameters are given in Table~\ref{linfit3mm}.
We tested the hypothesis that the slope $b$ is equal to 0 (which we took as the Null Hypothesis, $H_0$). 
In Table~\ref{linfit3mm} we show that, for the case of $b$, the $P$-values coming from the $t$-statistic are 62\%, 6\% and 29\% for $S_{\rm{3mm}}$, $p_{\rm{3mm}}$, and $\chi_{\rm{3mm}}$ respectively.
This points out that the probabilities of getting the fitted values of $b$ by chance are not small. 
Based on the values of the $t$-statistic, we also estimate the 95\% confidence intervals for the fitted values of $b$ (see Table~\ref{linfit3mm}). 
The zero-value (i.e. $b=0$) lies within all such 95\% confidence intervals.
Since the $P$-values are greater than the 5\% significance level the data do not provide grounds for rejecting the $H_0$ at 5\% significance level.
Therefore, the measurements of $S_{\rm{3mm}}$, $p_{\rm{3mm}}$, and $\chi_{\rm{3mm}}$ are each one of them distributed around a constant value, within a 95\% confidence level.

\input{./linfit-table-3mm} 

\input{./constantfit-table-3mm} 

Fitting a constant model (i.e., $y=a$) to the data further verifies the above picture (see also Fig~\ref{f3mm}). 
This is demonstrated by the remarkably low $P$-values for obtaining the fitted value by chance (see Table~\ref{constfit3mm}).
The resulting fitted constant of $S_{\rm{3mm}}$, $p_{\rm{3mm}}$, and $\chi_{\rm{3mm}}$ for \object{3C~286}, and their 95\% confidence intervals are  $S_{\rm{3mm}}=(0.91\pm0.02)$\,Jy, $p_{\rm{3mm}}=(13.5\pm0.3)$\,\%, and $\chi_{\rm{3mm}}=(37.3\pm0.8)^{\circ}$ (see Table~\ref{constfit3mm} and Fig.~\ref{f3mm}).
We propose these as the values of $S_{\rm{3mm}}$, $p_{\rm{3mm}}$, and $\chi_{\rm{3mm}}$ to use for calibration through observations of \object{3C~286}.

\subsection{1\,mm}
\label{res1mm}
At 1\,mm (Fig.~\ref{f1mm} and Table~\ref{t1mm}), the shorter observing bandwidth, the poorer atmospheric transmission and receiver sensitivity, and weakness of the source at this wavelength, together with the poorer time coverage at this spectral range in our monitoring of \object{3C~286} does not allow us to easily discern by visual inspection the stable $S_{\rm{1mm}}$, $p_{\rm{1mm}}$, and $\chi_{\rm{1mm}}$ pattern seen at 3\,mm.
However, for the linear fit of $S_{\rm{1mm}}(t)$, $p_{\rm{1mm}}(t)$, and $\chi_{\rm{1mm}}(t)$ (Table~\ref{linfit1mm}), the probabilities of getting the fitted values of $b$ by chance are also rather high; 99, 12 and 94\,\%, respectively.
The 95\% confidence intervals for the fitted values of $b$, for the cases of $S_{\rm{1mm}}(t)$, $p_{\rm{1mm}}(t)$, and $\chi_{\rm{1mm}}(t)$, also include $b=0$.
The corresponding $P$-values are greater than the 5\% significance level.
Thus, we can also confirm that the measurements of $S_{\rm{1mm}}$, $p_{\rm{1mm}}$, and $\chi_{\rm{1mm}}$ are distributed around a constant value, within a 95\% confidence level.  
This is confirmed by the constant fits to the 1\,mm data sets.
Their expected values, and 95\% confidence intervals, are $S_{\rm{1mm}}=(0.30\pm0.03)$\,\%, $p_{\rm{1mm}}=(14.4\pm1.8)$\,\%, and $\chi_{\rm{1mm}}=(33.1\pm5.7)^{\circ}$ (Table~\ref{constfit1mm} and Fig.~\ref{f1mm}).

\input{./linfit-table-1mm} 

\input{./constantfit-table-1mm} 

\subsection{Total flux and polarization properties of \object{3C~286} along the radio and millimeter spectrum}
\label{spectr}

\begin{figure}
   \centering
   \includegraphics[width=\columnwidth]{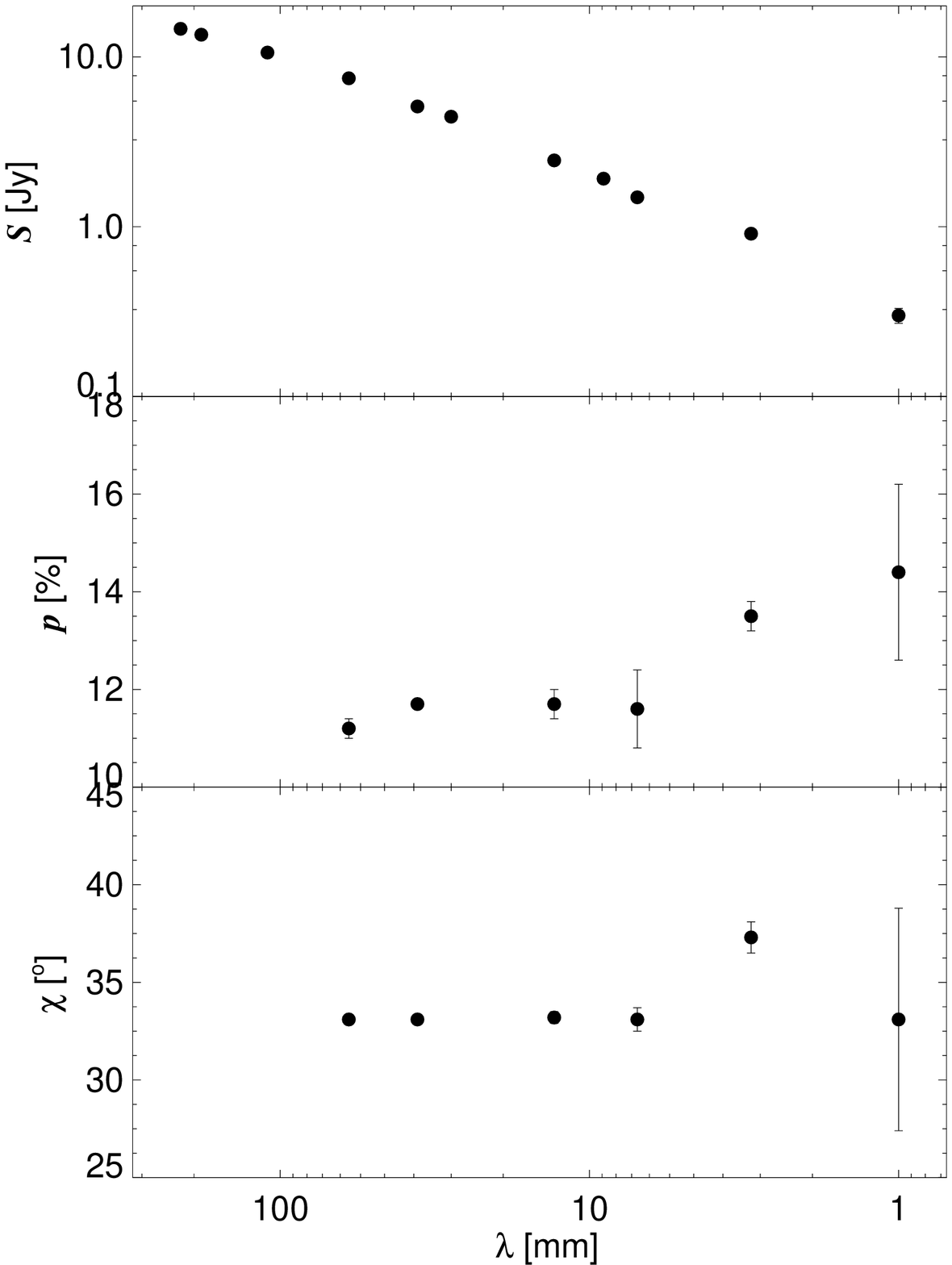}
   \caption{{Total flux density, polarization degree, and polarization angle of {3C~286} along the radio and millimeter spectrum. Plotted data are those presented in Table~\ref{spec-table}.}}
   \label{spec}
\end{figure}

Figure~\ref{spec} and Table~\ref{spec-table} show the total flux density and linear polarization properties of \object{3C~286} along the millimeter and radio-centimeter spectrum (from 1\,mm to 21\,cm).
The corresponding data comes from either the results presented on this paper, or those from \citet{Peng:2000p17257} and \citet{Taylor:2001p16867}, see Table~\ref{spec-table}.

\begin{figure}
   \centering
   \includegraphics[width=\columnwidth]{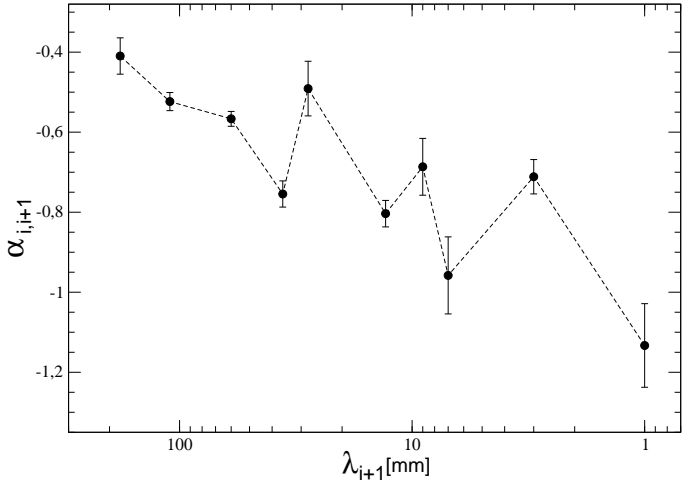}
   \caption{{Spectral index of 3C~286 as a function of $\nu_{\rm{i+1}}$. The data presented here are those in Table~\ref{spindx}.}}
   \label{spindex}
\end{figure}

The flux density plot shows a monotonically decreasing spectral trend as those expected for optically-thin synchrotron emission from relativistic jets \citep[e.g.,][]{1985ApJ...298..114M}.
Despite some moderated spectral-index jumps seen for the measurements with larger errors (Fig.~\ref{spindex} and Table~Ê\ref{spindx}), the spectral index also shows an overall decreasing pattern along the broad radio-millimeter spectral-range considered here.
No evident spectral break is observed from Fig.~\ref{spec}, hence suggesting that the synchrotron radiation observed from \object{3C~286} at 1\,mm through 21\,cm was produced by a single electron population.

The 3 and 1\,mm polarization properties of \object{3C~286} shown by Fig.~\ref{spec} seem to be rather similar to those at longer wavelengths within a $\sim2$\,\% fractional polarization, and $\sim4^{\circ}$ polarization angle.
However, the 3 and 1\,mm polarization degree ($p_{\rm{3mm}}\approx13.5$\,\% and $p_{\rm{1mm}}\approx14.5$\,\%, respectively) seem to increase slightly for decreasing wavelengths with regard to the fractional polarization at wavelengths longer than 3\,mm (that lies at the $[11,12]$\,\% level).
The polarization angle that we measure at 3\,mm ($\chi_{\rm{3mm}}\approx37^{\circ}$) is also moderately different than the one measured at longer wavelengths ($\sim33^{\circ}$, see Fig.~\ref{spec} and Table~\ref{spec-table}).
Our 1\,mm polarization angle measurement ($\chi_{\rm{1mm}}=33\pm6^{\circ}$) has sufficiently large uncertainty to avoid discerning if the shorter wavelength polarization angle matches the values of the 3\,mm measurements or those at longer wavelengths.
We hope to solve this ambiguity by better constraining the 1\,mm polarization angle of \object{3C~286} through the continued monitoring of the source that we currently perform at the IRAM 30\,m Telescope.

The slightly different fractional polarization and polarization angle at the shorter wavelengths in \object{3C~286} can be explained if the fraction of tangled magnetic field decreases towards inner regions upstream in the jet.
Some fraction of such tangled magnetic field is needed to explain the decreased fractional polarization observed in the entire radio and millimeter spectrum with regard to the predicted value for optically-thin polarized synchrotron-radiation \citep[$\sim70$\,\%,][]{Pacholczyk:1970}, if Faraday depolarization is negligible (which is clearly the case of \object{3C~286}, see Fig.~\ref{spec}).
Also, the bulk jet emission should be radiated from jet regions at inner upstream locations for shorter wavelengths, as usually observed through very long baseline interferometry in relativistic jets in active galactic nuclei \citep[e.g.,][]{Sokolovsky:2011p16166}.
Within this scenario, the radiation emitted at shorter wavelengths from inner jet region, would show higher fractional polarization than the longer wavelength emission.
The different polarization angles for radiation emitted from inner jet regions may be explained by a slightly different jet direction with regard to the jet section where the longer wavelength emission is radiated \citep[e.g.,][]{Abdo:2010p11811}.
This can be easily accommodated within Cotton et~al.'s model for \object{3C~286}  \citep[see Fig.~6 in][]{Cotton:1997p17000}, which uses a bend in the inner jet regions to explain the apparent lack of a high brightness temperature core --beamed away from Earth in their model-- as the cause for the unusual total flux and polarization properties of the source.

\input{./3C286-spectral-table} 

\input{./3C286-spindx-table} 

\section{Summary and conclusions}
\label{sum}
We have shown that although \object{3C~286} displays moderate millimeter flux densities ($S_{\rm{3mm}}\approx1$\,Jy and $S_{\rm{1mm}}\approx0.3$\,Jy), its large millimeter polarization degree ($p_{\rm{3mm}}\approx13.5$\,\% and $p_{\rm{1mm}}\approx14.5$\,\%) and its time stability make this source a valuable polarization calibrator for millimeter observations with instruments sensitive enough to detect linear polarization as bright as $\sim120$\,mJy at 3\,mm and $\sim45$\,mJy at 1\,mm.

The 3 and 1\,mm total flux and polarization properties of \object{3C~286} have been shown to be rather similar to those previously known at longer wavelengths, which further supports the suitability of extending the use of \object{3C~286} as total flux and polarization calibrator from radio towards short millimeter wavelengths.
The slightly larger 3 and 1\,mm fractional polarization of the source with regard to those at longer wavelengths ($p\in[11,12]$\,\% from 7\,mm to 6\,cm), can be explained if the degree of tangled magnetic field in the jet decreases towards inner upstream jet regions.
To explain the deviation of $\sim4^{\circ}$ of the 3\,mm (and perhaps 1\,mm) polarization angle with regard to that at longer wavelengths, a small bend in the inner jet regions is also needed, but this is fully consistent with previous models for \object{3C~286}.
The time-dependent behavior of the source at 3 and 1\,mm has been shown to be as stable as it is well known to be at shorter millimeter and radio wavelengths, which also supports the idea of the suitability of \object{3C~286} for total flux and polarization calibration purposes.

Having \object{3C~286} as an additional polarization calibrator may be an advantage for calibration and/or observing programs with limited observing time, if no other reliable polarization calibrator lies in the sky during their observing window.
\object{3C~286} has a maximum extension of up to $\sim3.5^{\prime\prime}$ on the plane of the sky.
In contrast, other available polarization calibrators show extended structures as large as $\sim5^{\prime}$, e.g. the case of the Crab nebula \citep{Aumont:2010p12769}, or even $\sim12^{\prime}$ for Cygnus A \citep{Zemcov:2010p16834}.
Hence, \object{3C~286} may facilitate polarization calibration observations with high resolution interferometers such as ALMA, Plateau de Bure, or the Sub-mm Array on extended configurations.
Moreover, given the moderate northern declination of \object{3C~286} (see Section~\ref{intr}) it can be observed comfortably from all northern millimeter observatories, and (at elevations up to $\approx30^{\circ}$) even from most southern millimeter observatories, including ALMA. 

We conclude that \object{3C~286} may be safely used for calibration of both single dish and interferometric polarization observations at 3\,mm, and possibly at shorter wavelengths.

\begin{acknowledgements}
The authors thank the anonymous referee for constructive revision of this paper.
The observations with the IRAM 30\,m Telescope were made remotely from many different places. 
We thank the Observatory staff for their competent support.
The IRAM 30m Telescope is supported by INSU/CNRS (France), MPG (Germany), and IGN (Spain).
This research has made use of the VLA/VLBA Polarization Calibration Database that was maintained by the National Radio Astronomy Observatory (NRAO) of the USA, until the end of 2009.
The NRAO is a facility of the USA's National Science Foundation operated under cooperative agreement by USA's Associated Universities, Inc.
This research was partly funded by the ``Consejer\'ia de Econom\'ia, Innovaci\'on y Ciencia'' of the Regional Government of Andaluc\'ia through grant P09-FQM-4784, an by the ``Ministerio de Econom\'ia y Competitividad'' of Spain through grant AYA2010-14844. DE acknowledges the Science and Technology Facilities Council (STFC) of the United Kingdom for support under grant ST/G003084/1.
\end{acknowledgements}


\end{document}

%% file: 3C286-table-3mm.tex
\begin{table}
\centering
\caption[]{3\,mm photo-polarimetric data from 3C~286.}
\begin{flushleft}
\begin{tabular} {lcccc}
\hline\noalign{\smallskip}
Date & RJD   &       $S_{\rm{3mm}}$     &     $p_{\rm{3mm}}$    &     $\chi_{\rm{3mm}}$  \\
     & days  &       Jy      &     \%     &   $^{\circ}$ \\
\hline\noalign{\smallskip}
2006-09-24 & 54003 & $ 0.92\pm 0.05$ & $15.6\pm0.5$ & $  39.0\pm 1.0$ \\
2006-10-03 & 54012 & $ 1.03\pm 0.05$ & $13.4\pm0.4$ & $  40.0\pm 1.1$ \\
2006-10-13 & 54022 & $ 0.90\pm 0.05$ & $12.2\pm0.4$ & $  39.0\pm 1.6$ \\
2006-10-14 & 54023 & $ 0.94\pm 0.05$ & $14.4\pm2.1$ & $  40.0\pm 2.6$ \\
2006-11-16 & 54056 & $ 0.87\pm 0.04$ & $12.8\pm1.1$ & $  32.1\pm 1.9$ \\
2006-11-17 & 54057 & $ 0.85\pm 0.04$ & $16.6\pm0.6$ & $  39.5\pm 0.9$ \\
2006-11-22 & 54062 & $ 0.90\pm 0.04$ & $13.9\pm0.5$ & $  41.8\pm 2.2$ \\
2006-12-01 & 54071 & $ 0.92\pm 0.05$ & $14.1\pm0.8$ & $  37.1\pm 1.1$ \\
2007-01-03 & 54104 & $ 0.96\pm 0.05$ & $14.0\pm1.4$ & $  38.4\pm 1.5$ \\
2007-01-17 & 54119 & $ 0.97\pm 0.05$ & $14.9\pm0.4$ & $  41.9\pm 1.0$ \\
2007-01-29 & 54130 & $ 0.94\pm 0.05$ & $13.4\pm0.4$ & $  36.1\pm 1.0$ \\
2007-06-11 & 54262 & $ 0.83\pm 0.04$ & $14.8\pm0.7$ & $  31.1\pm 1.3$ \\
2007-08-07 & 54320 & $ 1.00\pm 0.05$ & ... & ...  \\
2007-10-09 & 54383 & $ 0.85\pm 0.04$ & $15.6\pm0.4$ & $  37.4\pm 1.0$ \\
2008-05-04 & 54591 & $ 0.87\pm 0.06$ & $13.3\pm0.8$ & $  30.1\pm 1.9$ \\
2008-05-13 & 54600 & $ 0.95\pm 0.05$ & ... & ...  \\
2008-12-29 & 54830 & $ 1.02\pm 0.05$ & $13.5\pm0.4$ & $  39.1\pm 1.1$  \\
2009-06-18 & 55001 & $ 0.87\pm 0.04$ & $13.0\pm0.3$ & $  39.4\pm 0.8$  \\
2009-07-30 & 55043 & $ 0.90\pm 0.04$ & $14.9\pm1.1$ & $  38.9\pm 1.3$  \\
2009-08-18 & 55062 & $ 0.90\pm 0.04$ & $13.9\pm0.4$ & $  36.6\pm 2.1$  \\
2010-05-05 & 55322 & $ 0.98\pm 0.05$ & $13.6\pm0.4$ & $  37.5\pm 2.4$  \\
2010-06-06 & 55354 & $ 0.94\pm 0.05$ & $12.6\pm0.3$ & $  37.3\pm 1.1$  \\
2010-06-16 & 55364 & $ 0.83\pm 0.04$ & $13.5\pm0.3$ & $  37.5\pm 0.7$  \\
2010-08-15 & 55424 & $ 0.83\pm 0.04$ & $15.4\pm0.3$ & $  38.4\pm 0.7$  \\
2010-10-12 & 55482 & $ 0.93\pm 0.05$ & $13.0\pm0.3$ & $  33.8\pm 0.8$  \\
2011-01-31 & 55593 & $ 0.87\pm 0.04$ & $13.4\pm0.3$ & $  33.0\pm 0.7$  \\
2011-02-07 & 55600 & $ 0.95\pm 0.05$ & $13.5\pm0.3$ & $  35.2\pm 0.7$  \\
2011-03-25 & 55646 & $ 0.94\pm 0.05$ & $13.5\pm0.2$ & $  38.5\pm 0.7$  \\
2011-04-10 & 55662 & $ 0.92\pm 0.05$ & $13.4\pm0.3$ & $  37.6\pm 0.7$  \\
2011-04-25 & 55677 & $ 0.90\pm 0.04$ & $12.1\pm0.3$ & $  39.4\pm 0.9$  \\
2011-05-04 & 55686 & $ 0.96\pm 0.05$ & $13.2\pm0.3$ & $  35.7\pm 0.7$  \\
2011-05-22 & 55704 & $ 0.89\pm 0.04$ & $14.2\pm0.3$ & $  36.2\pm 0.7$  \\
2011-05-24 & 55706 & $ 0.98\pm 0.05$ & $15.1\pm0.3$ & $  41.1\pm 0.6$  \\
2011-06-15 & 55728 & $ 0.74\pm 0.04$ & $13.1\pm0.3$ & $  37.9\pm 0.8$  \\
2011-09-04 & 55809 & $ 0.82\pm 0.04$ & $13.6\pm0.3$ & $  35.9\pm 0.8$  \\
2011-09-11 & 55816 & $ 0.87\pm 0.04$ & $12.6\pm0.3$ & $  40.0\pm 0.8$  \\
2011-10-11 & 55846 & $ 0.95\pm 0.05$ & $12.2\pm0.3$ & $  38.0\pm 0.8$  \\
2011-10-16 & 55882 & $ 0.95\pm 0.05$ & $13.8\pm0.3$ & $  34.5\pm 0.7$ \\
2012-01-03 & 55930 & $ 0.97\pm 0.05$ & $13.3\pm0.3$ & $  37.4\pm 0.7$ \\
2012-01-30 & 55957 & $ 1.02\pm 0.05$ & $13.2\pm0.2$ & $  37.4\pm 0.7$ \\
\noalign{\smallskip}
\hline
\end{tabular}
\end{flushleft}
\label{t3mm}
\end{table}

%% file: 3C286-table-1mm.tex
\begin{table}
\centering
\caption[]{1\,mm photo-polarimetric data from 3C~286.}
\begin{flushleft}
\begin{tabular} {lcccc}
\hline\noalign{\smallskip}
Date & RJD   &       $S_{\rm{1mm}}$     &     $p_{\rm{1mm}}$    &     $\chi_{\rm{1mm}}$  \\
     & days  &       Jy      &     \%     &   $^{\circ}$ \\
\hline\noalign{\smallskip}
2010-05-05 & 55322 & $ 0.40\pm 0.02$ & $14.3\pm2.1$ & $  38.6\pm 3.9$  \\
2010-06-06 & 55354 & $ 0.33\pm 0.02$ & ... & ...  \\
2010-10-12 & 55482 & $ 0.34\pm 0.03$ & ... & ...  \\
2011-01-31 & 55593 & $ 0.31\pm 0.02$ & $18.0\pm2.4$ & $  22.3\pm 3.9$  \\
2011-02-07 & 55600 & $ 0.34\pm 0.02$ & $13.9\pm2.5$ & $  30.8\pm 5.0$  \\
2011-03-25 & 55646 & $ 0.28\pm 0.02$ & $17.7\pm3.4$ & $  35.8\pm 6.2$  \\
2011-04-10 & 55662 & $ 0.23\pm 0.01$ & $17.5\pm3.2$ & $  23.7\pm 5.2$  \\
2011-04-25 & 55677 & $ 0.21\pm 0.02$ & $13.4\pm5.5$ & $  19.3\pm11.9$  \\
2011-05-04 & 55686 & $ 0.28\pm 0.02$ & $18.5\pm2.8$ & $  41.6\pm 4.0$  \\
2011-09-11 & 55816 & $ 0.26\pm 0.02$ & $13.9\pm4.7$ & $  38.1\pm 9.7$  \\
2011-10-11 & 55846 & $ 0.31\pm 0.02$ & $17.0\pm5.8$ & $  47.9\pm 8.6$  \\
2011-10-16 & 55882 & $ 0.34\pm 0.02$ & $14.1\pm2.5$ & $  32.2\pm 5.1$ \\
2012-01-03 & 55930 & $ 0.39\pm 0.02$ & $10.2\pm1.8$ & $  47.1\pm5.4$ \\
2012-01-30 & 55957 & $ 0.36\pm 0.02$ & $13.2\pm2.1$ & $  23.8\pm 4.6$ \\
\noalign{\smallskip}
\hline
\end{tabular}
\end{flushleft}
\label{t1mm}
\end{table}

%% file: linfit-table-3mm.tex
\begin{table*}
\centering
\caption[]{{Best-fit linear model parameters for $S_{\rm{3mm}}(t)$, $p_{\rm{3mm}}(t)$, and $\chi_{\rm{3mm}}(t)$.}}
\begin{flushleft}
\begin{tabular} {ccccccc}
\hline\noalign{\smallskip}
Par. & Est.   &     $\epsilon$    &    $t$-stat    &   $P$ & \multicolumn{2}{c}{95\% Conf. Int.} \\
(1) & (2)   &   (3)   &  (4)   &  (5) & (6) & (7) \\
\hline\noalign{\smallskip}
\multicolumn{7}{c}{$S_{\rm{3mm}}$}\\
\hline\noalign{\smallskip}
$a$ & 0.915 &  0.018 & 51 & $1.1\times10^{-36}$ & 0.879 & 0.951 \\
$b$ & $-7\times10^{-6}$ & $1.4\times10^{-5}$ &  -0.5 &  0.62 & $-4\times10^{-5}$ & $2.2\times10^{-5}$\\
\hline\noalign{\smallskip}
\multicolumn{7}{c}{$p_{\rm{3mm}}$}\\
\hline\noalign{\smallskip}
$a$ & 14.18 &  0.36 &  39 &  $4\times10^{-31}$ &  13.45 & 14.91 \\
$b$ & $-5\times10^{-4}$ &  $2.4\times10^{-4}$ &  -1.9 &  0.06 &  $-9\times10^{-4}$ & $3\times10^{-5}$\\
\hline\noalign{\smallskip}
\multicolumn{7}{c}{$\chi_{\rm{3mm}}$}\\
\hline\noalign{\smallskip}
$a$ & 38.19 &  0.88 &  43 &  $1.1\times10^{-32}$ &  36.41 & 39.98\\
$b$ & $-6\times10^{-4}$ &  $6\times10^{-4}$ &  -1.07 &  0.29 & $-1.8\times10^{-3}$ & $6\times10^{-4}$\\
\noalign{\smallskip}
\hline
\end{tabular}
\end{flushleft}
{Note: Columns are as follows: (1) fitted parameter, (2) estimate from the fit, (3) standard error, (4) $t$-statistic, (5) its corresponding $P$-value, and (6) and (7) lower and upper limits of the 95\% confidence interval, respectively.}
\label{linfit3mm}
\end{table*}

%% file: constantfit-table-3mm.tex
\begin{table*}
\centering
\caption[]{{{Best-fit constant model parameters for $S_{\rm{3mm}}(t)$, $p_{\rm{3mm}}(t)$, and $\chi_{\rm{3mm}}(t)$.}}}
\begin{flushleft}
\begin{tabular} {cccccc}
\hline\noalign{\smallskip}
$a$   &     $\epsilon$    &    $t$-stat    &   $P$ & \multicolumn{2}{c}{95\% Conf. Int.} \\
(1)   &   (2)   &  (3)   &  (4) & (5) & (6) \\
\hline\noalign{\smallskip}
\multicolumn{6}{c}{$S_{\rm{3mm}}$}\\
\hline\noalign{\smallskip}
0.908  & 0.010  &      89 &     $1.2\times10^{-46}$ &   0.887 & 0.928 \\
\hline\noalign{\smallskip}
\multicolumn{6}{c}{$p_{\rm{3mm}}$}\\
\hline\noalign{\smallskip}
13.54   &  0.15      &   93     &  $2.2\times10^{-45}$  &  13.25 & 13.84 \\
\hline\noalign{\smallskip}
\multicolumn{6}{c}{$\chi_{\rm{3mm}}$}\\
\hline\noalign{\smallskip}
37.35  &  0.38       &  97   &     $3\times10^{-46}$ &  36.57 & 38.13 \\
\noalign{\smallskip}
\hline
\end{tabular}
\end{flushleft}
{Note: Columns are as follows: (1) estimate of $a$ from the fit, (2) standard error, (3) $t$-statistic, (4) its corresponding $P$-value, and (5) and (6) lower and upper limits of the 95\% confidence interval, respectively.}
\label{constfit3mm}
\end{table*}

%% file: linfit-table-1mm.tex
\begin{table*}
\centering
\caption[]{{Same as Table~\ref{linfit3mm} but for the 1\,mm data.}}
\begin{flushleft}
\begin{tabular} {ccccccc}
\hline\noalign{\smallskip}
Par. & Est.   &     $\epsilon$    &    $t$-stat    &   $P$ & \multicolumn{2}{c}{95\% Conf. Int.} \\
(1) & (2)   &   (3)   &  (4)   &  (5) & (6) & (7) \\
\hline\noalign{\smallskip}
\multicolumn{7}{c}{$S_{\rm{1mm}}$}\\
\hline\noalign{\smallskip}
$a$ & 0.303 & 0.120 & 2.5 & $2.4\times10^{-2}$ & 0.046 & 0.560 \\
$b$ & $-8\times10^{-7}$ & $8\times10^{-5}$ & -0.009 & 0.99 & $-1.8\times10^{-4}$ & $1.8\times10^{-4}$ \\
\hline\noalign{\smallskip}
\multicolumn{7}{c}{$p_{\rm{1mm}}$}\\
\hline\noalign{\smallskip}
$a$ & 16.84 & 1.61 & 10 & $1.1\times10^{-6}$ & 13.25 & 20.43 \\
$b$ & $-6\times10^{-3}$ & $4\times10^{-3}$ & -1.7 & 0.12 & $-1.4\times10^{-2}$ & $2.0\times10^{-3}$ \\
\hline\noalign{\smallskip}
\multicolumn{7}{c}{$\chi_{\rm{1mm}}$}\\
\hline\noalign{\smallskip}
$a$ & 33.46 & 5.69 & 6 & $1.6\times10^{-4}$ & 20.78 & 46.13 \\
$b$ & $-1\times10^{-3}$ &  0.014 & -0.08 & 0.94 & -0.03 & 0.03 \\
\noalign{\smallskip}
\hline
\end{tabular}
\end{flushleft}
\label{linfit1mm}
\end{table*}

%% file: constantfit-table-1mm.tex
\begin{table*}
\centering
\caption[]{{{Same as Table~\ref{constfit3mm} but for the 1\,mm data.}}}
\begin{flushleft}
\begin{tabular} {cccccc}
\hline\noalign{\smallskip}
$a$   &     $\epsilon$    &    $t$-stat    &   $P$ & \multicolumn{2}{c}{95\% Conf. Int.} \\
(1)   &   (2)   &  (3)   &  (4) & (5) & (6) \\
\hline\noalign{\smallskip}
\multicolumn{6}{c}{$S_{\rm{1mm}}$}\\
\hline\noalign{\smallskip}
0.302  & 0.014 & 21 & $1.2\times10^{-12}$ &  0.272   &  0.332 \\
\hline\noalign{\smallskip}
\multicolumn{6}{c}{$p_{\rm{1mm}}$}\\
\hline\noalign{\smallskip}
14.44  &  0.82 & 18 &  $2.0\times10^{-9}$  &  12.64  & 16.24 \\
\hline\noalign{\smallskip}
\multicolumn{6}{c}{$\chi_{\rm{1mm}}$}\\
\hline\noalign{\smallskip}
33.07  &  2.60  &  13  &  $6\times10^{-8}$ &  27.35  &  38.80 \\
\noalign{\smallskip}
\hline
\end{tabular}
\end{flushleft}
\label{constfit1mm}
\end{table*}

%% file: 3C286-spectral-table.tex
\begin{table}
\centering
\caption[]{{$S$, $p$, and $\chi$ measurements of 3C~286 along the millimeter and radio spectrum.}}
\begin{flushleft}
\begin{tabular} {lcccc}
\hline\noalign{\smallskip}
 $\lambda$   &       $S$     &     $p$    &     $\chi$   & Ref.\\
  mm   &       Jy      &     \%     &   $^{\circ}$ &     \\
\hline\noalign{\smallskip}
1\tablefootmark{a} & $ 0.30\pm 0.03$ & $14.4\pm1.8$ & $  33.1\pm 5.7$  & 1 \\
3\tablefootmark{a}  & $ 0.91\pm 0.02$ & $13.5\pm0.3$ & $  37.3\pm 0.8$  & 1 \\
7         & $ 1.49\pm 0.03$\tablefootmark{b} & $11.6\pm0.8$\tablefootmark{c} & $  33.1\pm 0.6$\tablefootmark{c}  & 2,3 \\
9         & $ 1.92\pm 0.03$\tablefootmark{b} & ... & ...  & 2 \\
13         & $ 2.46\pm 0.05$\tablefootmark{b} & $11.7\pm0.3$\tablefootmark{c} & $  33.2\pm 0.3$\tablefootmark{c}  & 2,3 \\
28         & $ 4.45\pm 0.06$\tablefootmark{b} & ... & ...  & 2 \\
36        & $ 5.11\pm 0.07$\tablefootmark{b} & $11.7\pm0.1$\tablefootmark{c} & $  33.1\pm 0.1$\tablefootmark{c}  & 2,3 \\
60         & $ 7.49\pm 0.07$\tablefootmark{b} & $11.2\pm0.2$\tablefootmark{c} & $  33.1\pm 0.2$\tablefootmark{c}  & 2,3 \\
110        & $10.62\pm 0.07$\tablefootmark{b} & ... & ...  & 2 \\
180       & $13.53\pm 0.11$\tablefootmark{b} & ... & ...  & 2 \\
210       & $14.65\pm 0.05$\tablefootmark{b} & ... & ...  & 2 \\
\noalign{\smallskip}
\hline
\end{tabular}
\end{flushleft}
\label{spec-table}
\tablefoot{
\tablefoottext{a}{{$S$, $p$, and $\chi$ correspond to fitted values as presented in this work.}}
\tablefoottext{b}{{Average values of $S$ are those presented by \citet{Peng:2000p17257}.}}
\tablefoottext{c}{{Average values of measurements made during 2009 by \citet{Taylor:2001p16867}, see {\tt http://www.vla.nrao.edu/astro/calib/polar/2009/}.}}
}
\tablebib{{(1) This paper; (2) \citet{Peng:2000p17257}; (3) \citet{Taylor:2001p16867}}}

\end{table}

%% file: 3C286-spindx-table.tex
\begin{table}
\centering
\caption[]{{Spectral indexes of flux densities for contiguous wavelenghts as shown in Table~\ref{spec-table}.}}
\begin{flushleft}
\begin{tabular} {ccccc}
\hline\noalign{\smallskip}
 $\lambda_{\rm{i+1}}$   &  $\nu_{\rm{i+1}}$&      $\lambda_{\rm{i}}$ &  $\nu_{\rm{i+1}}$     &     $\alpha_{\rm{i,i+1}}$\\
mm &  GHz   & mm &     GHz    &     \\
\hline\noalign{\smallskip}
1    & 229 & 3    &  86  & $-1.13\pm0.11$ \\
3    & 86   & 7    &  43  & $-0.71\pm0.04$ \\
7    & 43   & 9    &  33  & $-0.96\pm0.10$ \\
9    & 33   & 13  &  22  & $-0.69\pm0.07$ \\
13  & 22   & 28  &  11  & $-0.80\pm0.03$ \\
28  & 11   & 36  &  8.4 & $-0.49\pm0.07$ \\
36  & 8.4  & 60  &  5   & $-0.75\pm0.03$ \\
60  &  5    & 110  & 2.7 & $-0.57\pm0.02$ \\
110  & 2.7 & 180  & 1.7 & $-0.52\pm0.02$ \\
180  & 1.7  & 210 & 1.4 & $-0.41\pm0.05$ \\
\noalign{\smallskip}
\hline
\end{tabular}
\end{flushleft}
\label{spindx}
\tablefoot{{The spectral index is defined such that $\alpha_{\rm{i,i+1}}=\frac{log(S_{\rm{i}}/S_{\rm{i+1}})}{log(\nu_{\rm{i}}/\nu_{\rm{i+1}})}$.}}
\end{table}

%% file: 3C286-pol-calib-v7.0.bbl
\begin{thebibliography}{}

\bibitem[\protect\citeauthoryear{Abdo et~al.}{Abdo
  et~al.}{2010}]{Abdo:2010p11811}
  Abdo, A. A., Ackermann, M., Ajello, M., et~al., 2010, Nature, 463, 919

\bibitem[\protect\citeauthoryear{Agudo et~al.}{Agudo
  et~al.}{2006}]{Agudo:2006p203}
Agudo I., Krichbaum T.~P., Ungerechts H., et~al., 2006, A{\&}A, 456, 117

\bibitem[\protect\citeauthoryear{Agudo et~al.}{Agudo
  et~al.}{2010}]{Agudo:2010p12104}
Agudo I., Thum C., Wiesemeyer H.,  Krichbaum T.~P., 2010, ApJS, 189, 1

\bibitem[\protect\citeauthoryear{Agudo et~al.}{Agudo
  et~al.}{2011a}]{Agudo:2011p14707}
Agudo I., Jorstad S.~G., Marscher A.~P., et~al., 2011a, ApJL, 726, L13

\bibitem[\protect\citeauthoryear{Agudo et~al.}{Agudo
  et~al.}{2011b}]{Agudo:2011p15946}
Agudo I., Marscher A.~P., Jorstad S.~G., et~al., 2011b, ApJL, 735, L10

\bibitem[\protect\citeauthoryear{Akujor \& Garrington}{Akujor \&
  Garrington}{1995}]{Akujor:1995p16994}
Akujor C.~E.,  Garrington S.~T., 1995, A{\&}AS, 112, 235

\bibitem[\protect\citeauthoryear{Aumont et~al.}{Aumont
  et~al.}{2010}]{Aumont:2010p12769}
Aumont J., Conversi L., Thum C., et~al., 2010, A{\&}A, 514, 70

\bibitem[\protect\citeauthoryear{Baars et~al.}{Baars
  et~al.}{1977}]{Baars:1977p16842}
Baars J.~W.~M., Genzel R., Pauliny-Toth I.~I.~K.,  Witzel A., 1977, A{\&}A, 61,
  99

\bibitem[\protect\citeauthoryear{Barvainis, Clemens, \& Leach}{Barvainis
  et~al.}{1988}]{Barvainis:1988p16835}
Barvainis R., Clemens D.~P.,  Leach R., 1988, AJ, 95, 510

\bibitem[\protect\citeauthoryear{Burbidge \& Burbidge}{Burbidge \&
  Burbidge}{1969}]{Burbidge:1969p16881}
Burbidge G.~R.,  Burbidge E.~M., 1969, Nature, 224, 21

\bibitem[\protect\citeauthoryear{Cotton et~al.}{Cotton
  et~al.}{1997}]{Cotton:1997p17000}
Cotton W.~D., Fanti C., Fanti R., et~al., 1997, A{\&}A, 325, 479

\bibitem[\protect\citeauthoryear{Fanti et~al.}{Fanti
  et~al.}{1985}]{Fanti:1985p16888}
Fanti C., Fanti R., Parma P., Schilizzi R.~T.,  van Breugel W.~J.~M., 1985,
  A{\&}A, 143, 292

\bibitem[\protect\citeauthoryear{G{\'o}mez et~al.}{G{\'o}mez
  et~al.}{2002}]{2002.VLBA.SM.30}
G{\'o}mez J.~L., Marscher A.~P., Alberdi A., Jorstad S.~G.,  Agudo I., 2002,
  VLBA Scientific Memorandum, 30, 1
  (http://www.vlba.nrao.edu/memos/sci/sci30memo.ps)

\bibitem[\protect\citeauthoryear{Jiang et~al.}{Jiang
  et~al.}{1996}]{Jiang:1996p17007}
Jiang D.~R., Dallacasa D., Schilizzi R.~T., et~al., 1996, A{\&}A, 312, 380

\bibitem[\protect\citeauthoryear{Kraus et~al.}{Kraus
  et~al.}{1999}]{Kraus:1999p13882}
Kraus A., Quirrenbach A., Lobanov A.~P., et~al., 1999, A{\&}A, 344, 807

\bibitem[\protect\citeauthoryear{Marscher \& Gear}{Marscher \& Gear}
{1985}]{1985ApJ...298..114M}
Marscher A. P \& Gear, W. K., 1985, ApJ, 298, 114

\bibitem[\protect\citeauthoryear{Matsumura et~al.}{Matsumura
  et~al.}{2010}]{Matsumura:2010p16833}
Matsumura T., Ade P., Barkats D., et~al., 2010, Proc. of SPIE, 7741, 77412O

\bibitem[\protect\citeauthoryear{McKinnon}{McKinnon}{1992}]{McKinnon:1992p16861}
McKinnon M.~M., 1992, A{\&}A, 260, 533

\bibitem[\protect\citeauthoryear{Ott et~al.}{Ott et~al.}{1994}]{Ott:1994p16851}
Ott M., Witzel A., Quirrenbach A., et~al., 1994, A{\&}A, 284, 331

\bibitem[\protect\citeauthoryear{Pacholczyk}{Pacholczyk}{1970}]{Pacholczyk:1970}
Pacholczyk, A. G. 1970, Radio astrophysics. Nonthermal Processes in Galactic
and Extragalactic Sources (San Francisco: Freeman)

\bibitem[\protect\citeauthoryear{Peacock \& Wall}{Peacock \&
  Wall}{1982}]{Peacock:1982p16871}
Peacock J.~A.,  Wall J.~V., 1982, MNRAS, 198, 843

\bibitem[\protect\citeauthoryear{Peng et~al.}{Peng et~al.}{2000}]{Peng:2000p17257}
Peng, B., Kraus, A., Krichbaum, T. P., Witzel, A., 2000, A\&AS, 145 ,1

\bibitem[\protect\citeauthoryear{Perley}{Perley}{1982}]{Perley:1982p6054}
Perley R.~A., 1982, AJ, 87, 859

\bibitem[\protect\citeauthoryear{Rudnick \& Jones}{Rudnick \&
  Jones}{1983}]{Rudnick:1983p17021}
Rudnick L.,  Jones T.~W., 1983, AJ, 88, 518

\bibitem[\protect\citeauthoryear{Sokolovsky et~al.}{Sokolovsky 
et~al.}{2011}]{Sokolovsky:2011p16166}
Sokolovsky, K. V., Kovalev, Y. Y., Pushkarev, A. B.,  Lobanov, A. P.,  2011, A\&A, 532, A38

\bibitem[\protect\citeauthoryear{Taylor et~al.}{Taylor
  et~al.}{2001}]{Taylor:2001p16867}
Taylor , GB~, Myers ,  ST~, 2001, VLBA Scientific Memorandum, 26, 1
  (http://www.vlba.nrao.edu/memos/sci/sci26memo.ps)

\bibitem[\protect\citeauthoryear{Thum et~al.}{Thum
  et~al.}{2003}]{Thum:2003p17117}
Thum C., Wiesemeyer H., Morris D., Navarro S.,  Torres M., 2003, Proc. of SPIE,
  4843, 272

\bibitem[\protect\citeauthoryear{Thum et~al.}{Thum
  et~al.}{2008}]{2008PASP..120..777T}
Thum C., Wiesemeyer H., Paubert G., Navarro S.,  Morris D., 2008, PASP, 120,
  777

\bibitem[\protect\citeauthoryear{Zemcov et~al.}{Zemcov
  et~al.}{2010}]{Zemcov:2010p16834}
Zemcov M., Ade P., Bock J., et~al., 2010, ApJ, 710, 1541

\end{thebibliography}
